\documentclass[a4paper]{jpconf}
\usepackage{graphicx}
\usepackage{xcolor}
\usepackage{rotating}
\usepackage{listings}
\usepackage{multirow}
\usepackage[ampersand]{easylist}
\usepackage{hyperref}
\usepackage{comment}
\usepackage{bm}
\usepackage{subcaption}
\usepackage{lineno}
\begin{document}
\title{Heavy-flavour correlations and jets with ALICE at the LHC}

\author{Antonio Carlos Oliveira da Silva, for the ALICE Collaboration}

\address{Physics Department, University of Tennessee - Knoxville, USA}

\ead{antonio.silva@cern.ch}

\begin{abstract}

This contribution summarises the results on heavy-flavour correlations and jets measured with ALICE detector. Studies of D$^{0}$-tagged jets are presented in various collision systems and energies. The measurement of the fraction of jet momentum carried by the D$^{0}$ meson in the direction of the jet axis ($z_{\rm ||}^{\rm ch,jet}$) in pp collisions at $\sqrt{s}$ = 5.02 and 13 TeV as well as the nuclear modification factor in p--Pb and Pb--Pb collisions are presented. The measurements of the production of jets tagged by electrons from heavy-flavour decays in pp and p--Pb are compared to expectations from POWHEG+PYTHIA8 simulations. The $p_{\rm T}$-differential cross section of b-tagged jets in p--Pb collisions is also reported and compared to expectations from POWHEG event generator. The azimuthal correlation of D mesons with charged particles in pp and p--Pb collisions are also presented. 
\end{abstract}

\section{Introduction}
Heavy quarks are produced in hard scatterings in the first stages of hadronic collisions. Therefore, they experience the whole evolution of the quark-gluon plasma (QGP) in heavy-ion collisions. Furthermore, because of their relatively large mass, heavy-quark production is well described by perturbative QCD calculations. These features make heavy-quarks ideal probes of the hot medium.
Measurements of azimuthal correlations of D mesons with charged particles and heavy-flavour jets in pp collisions are a valuable test of pQCD-based models describing heavy-quark production and also serve as reference for measurements in heavy-ion collisions that provide information on cold and hot nuclear matter effects.
In this proceedings the latest results on heavy-flavour correlations and jets are presented. Jets in this contribution are reconstructed using anti-$k_{\rm t}$ algorithm as implemented in FASTJET \cite{Cacciari_2012}.

\section{Analysis methods and results}

\subsection{D$^{0}$-tagged jets}
\label{DJets}
The D$^{0}$ mesons are reconstructed through their hadronic decay channel D$^{0}\to\rm{ K^{-}\pi^{+}}$ (BR = 3.89\%) exploiting the excellent particle-identification capabilities of the ALICE detector \cite{Botta:2017bwj}. After replacing the 4-momentum vector of the D$^{0}$-decay daughter particles with the D$^{0}$ one, jets are reconstructed with resolution parameter $R$ = 0.3. The feed-down contribution from bottom quark decays is removed using POWHEG+PYTHIA6 \cite{Frixione_2007,Sj_strand_2006} simulations. A Bayesian unfolding \cite{Prosper:2011zz} is performed to correct the jet momentum for detector effects and fluctuations in the underlying-event energy density. Additional details on the method can be found in \cite{Acharya:2019zup}.
The momentum fraction of the D$^{0}$ with respect to the jet is defined as $z_{\rm ||}^{\rm ch,jet} = \frac{\vec{p}_{\rm D} . \vec{p}_{\rm ch,jet}}{||\vec{p}_{\rm ch,jet}||^{2}}$, where $\vec{p}_{D}$ and $\vec{p}_{ch,jet}$ are the momenta of the D$^{0}$ and the jet, respectively.

\begin{figure}[bth]
\centering
\begin{subfigure}[b]{0.48\textwidth}
\includegraphics[width=20pc]{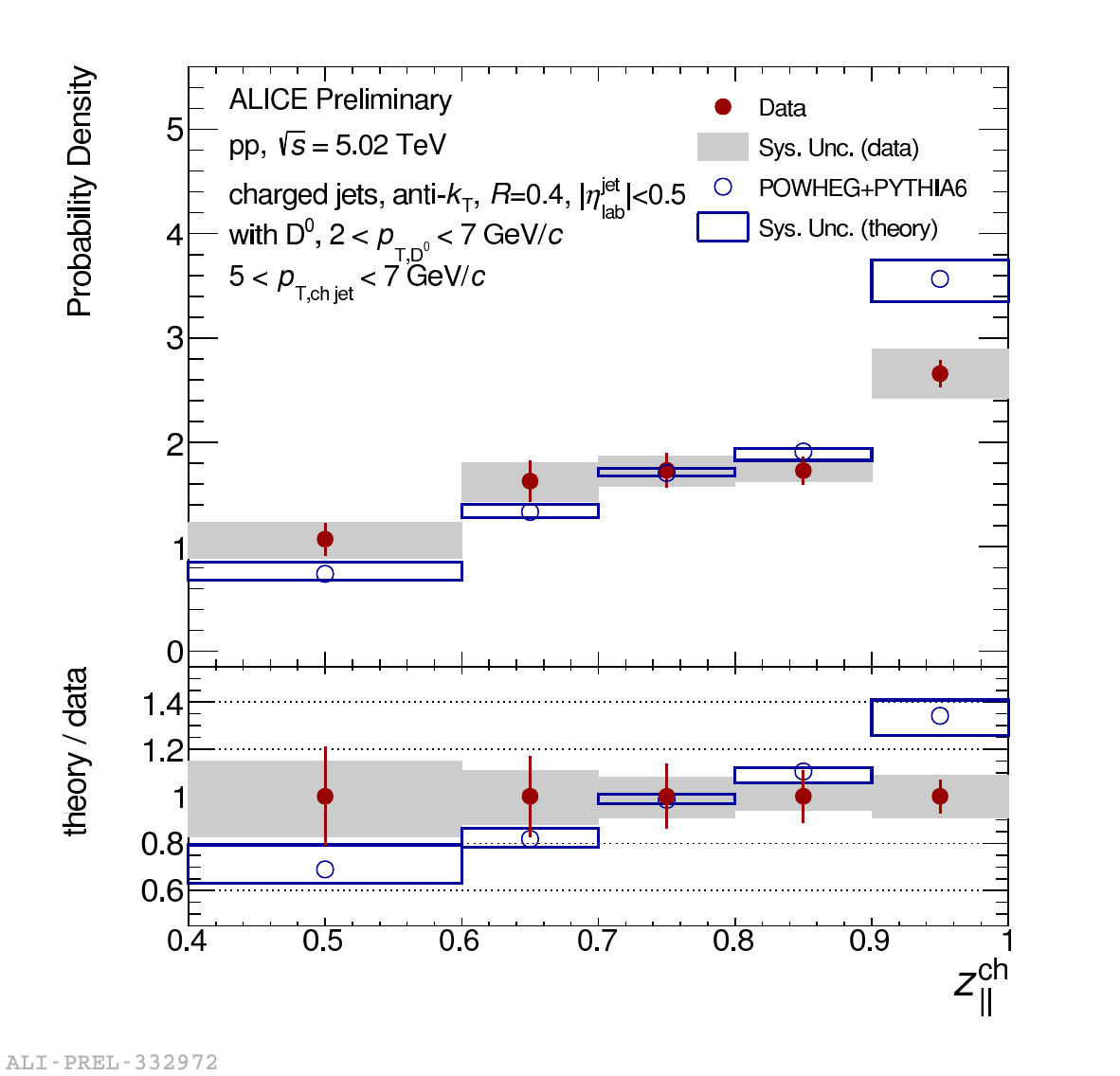}
\caption{Lower jet momentum region.}
\label{fZ5TeV_lowMom}
\end{subfigure}
\begin{subfigure}[b]{0.48\textwidth}
\includegraphics[width=20pc]{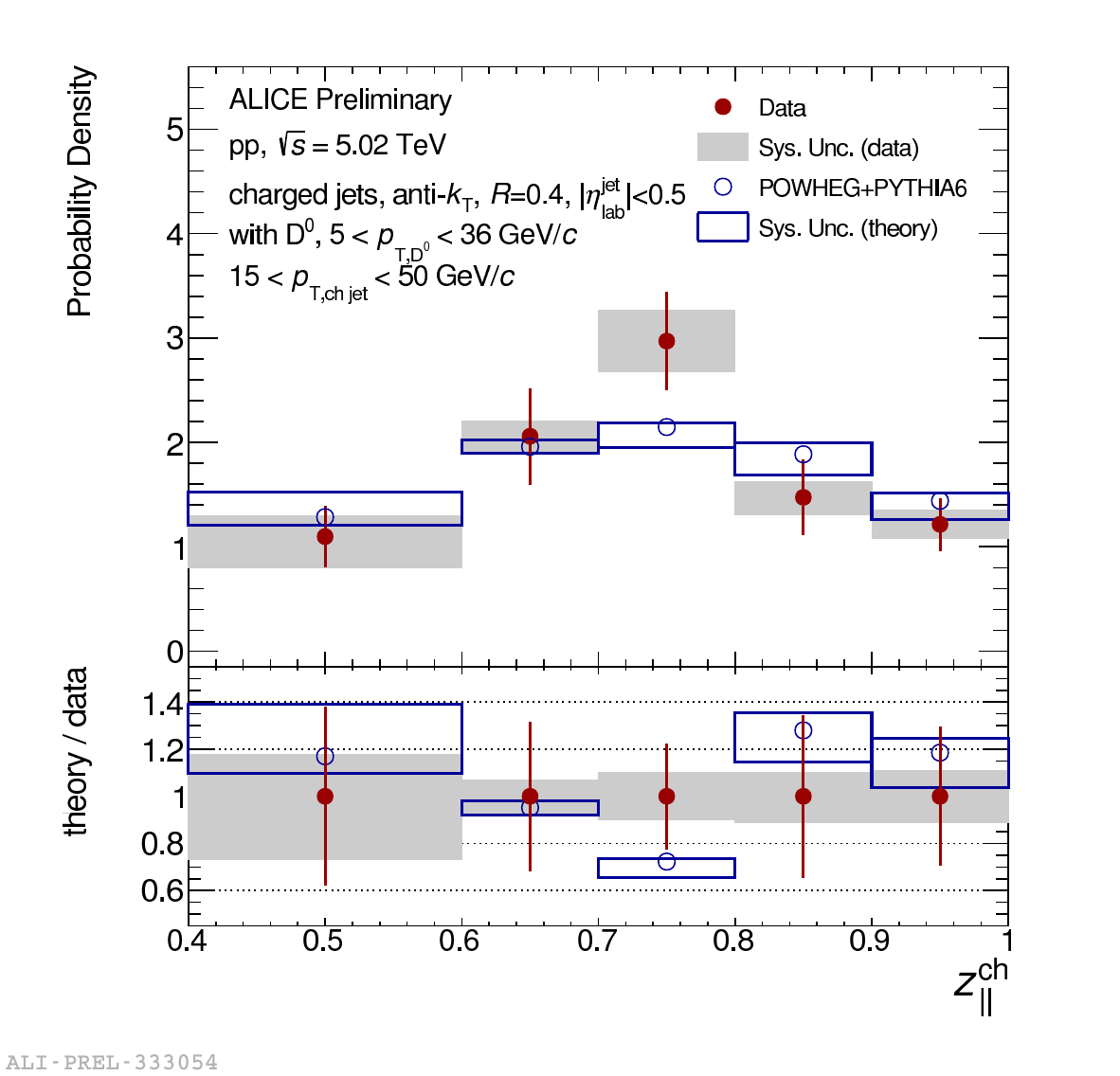}
\caption{Higher jet momentum region.}
\label{fZ5TeV_highMom}
\end{subfigure}
\caption{Momentum fraction of D$^{0}$-tagged jets in pp collisions at $\sqrt{s}=$ 5.02 TeV.}
\label{fZ5TeV}
\end{figure}

\begin{figure}[bth]
\centering
\begin{subfigure}[b]{0.48\textwidth}
\includegraphics[width=18pc]{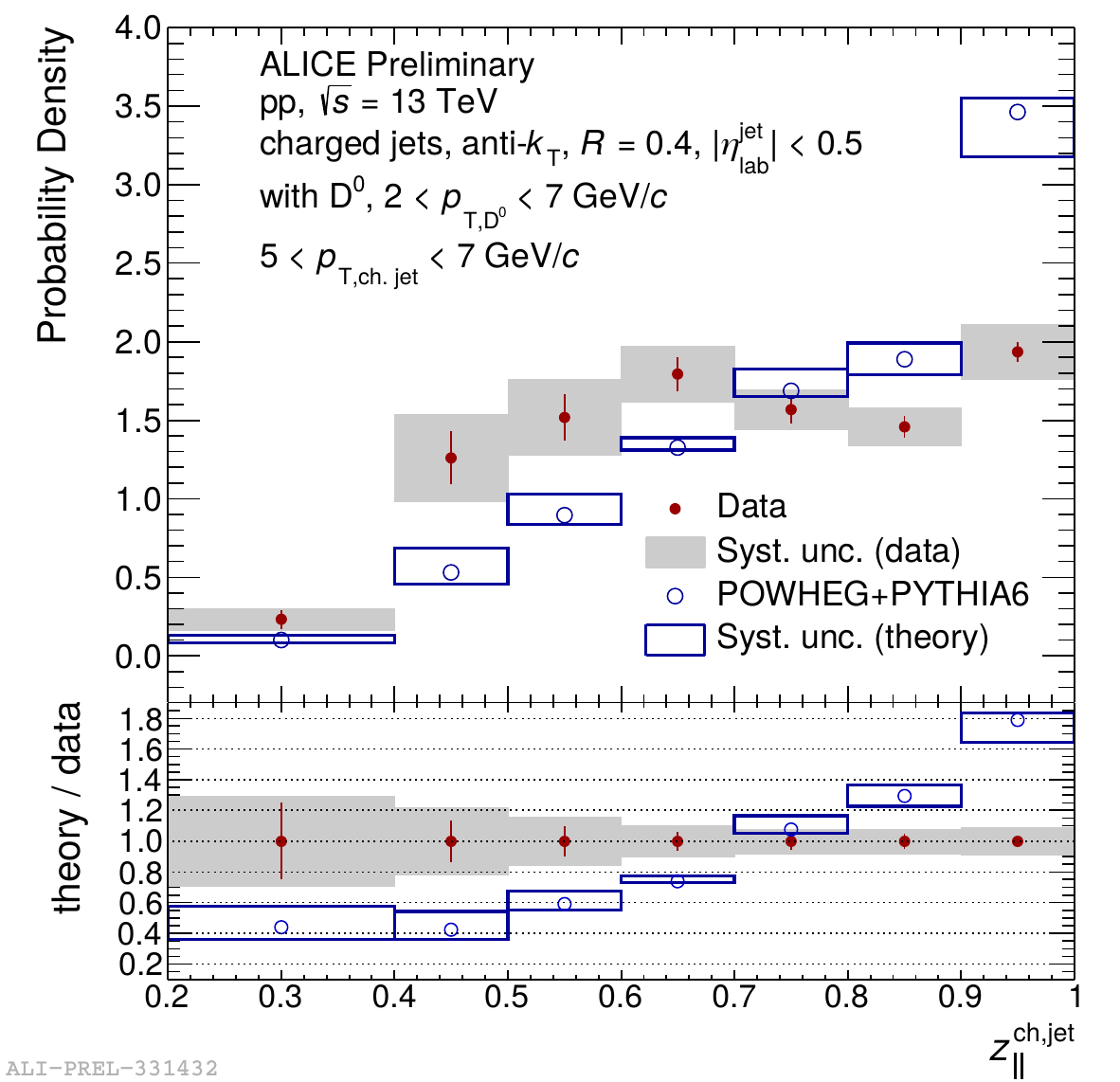}
\caption{Lower jet momentum region.}
\label{fZ13TeV_lowMom}
\end{subfigure}
\begin{subfigure}[b]{0.48\textwidth}
\includegraphics[width=18pc]{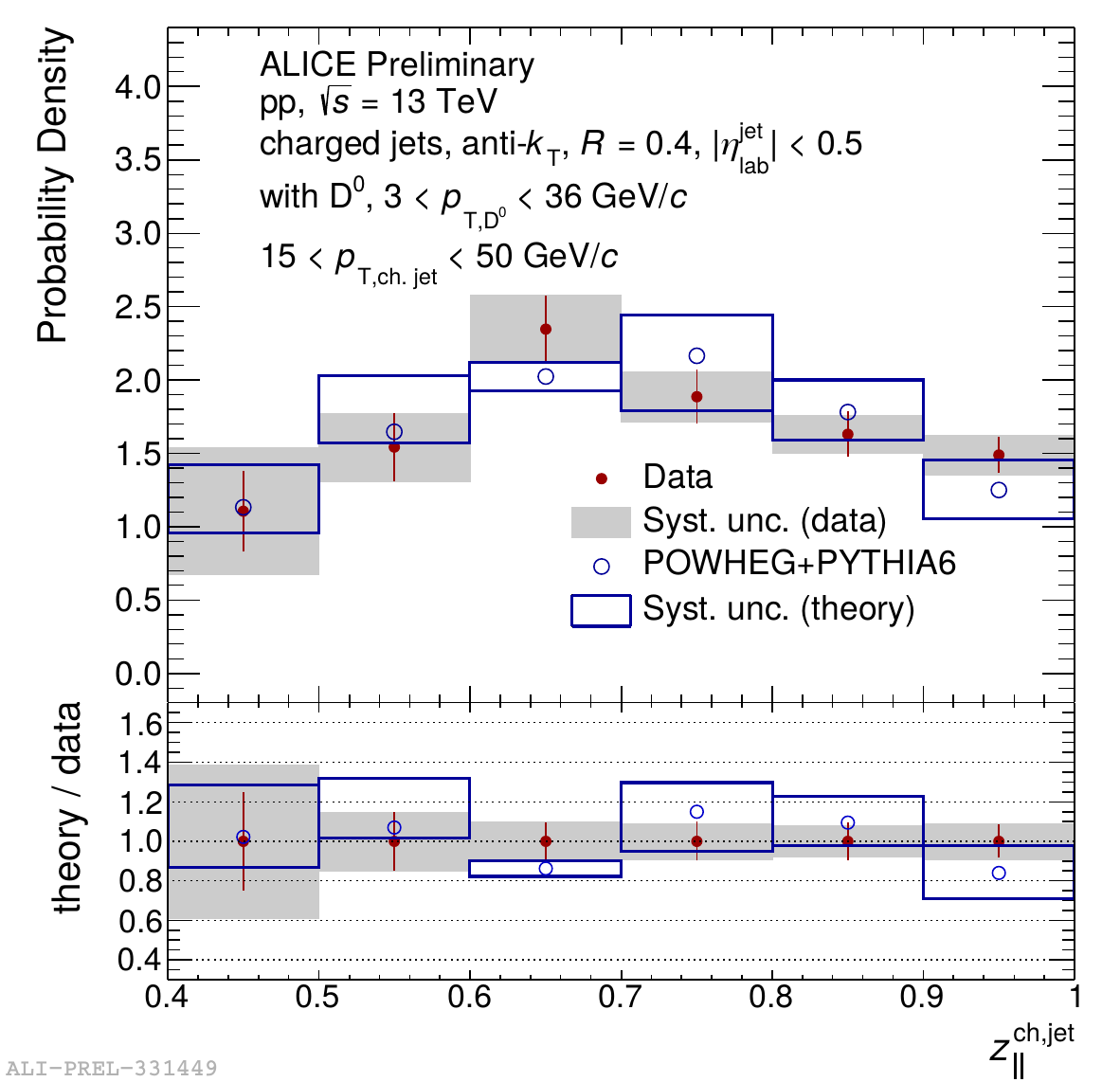}
\caption{Higher jet momentum region.}
\label{fZ13TeV_highMom}
\end{subfigure}
\caption{Momentum fraction of D$^{0}$-tagged jets in pp collisions at $\sqrt{s}=$ 13 TeV.}
\label{fZ13TeV}
\end{figure}

The $z_{\rm ||}^{\rm ch,jet}$ probability density for D$^{0}$-tagged jets in pp collisions at a center-of-mass energy $\sqrt{s}$ = 5.02 and 13 TeV are presented in Fig. \ref{fZ5TeV} and \ref{fZ13TeV}.

At low jet momentum, the measured probability densities favor large values of $z_{\rm ||}^{\rm ch,jet}$, indicating a hard fragmentation of the charm quark. The higher jet momentum region (Fig. \ref{fZ5TeV_highMom} and \ref{fZ13TeV_highMom}) presents a flatter $z_{\rm ||}^{\rm ch,jet}$ distribution in comparison to the lower momentum region. When compared with POWHEG+PYTHIA6 simulations, the data of the 5 $<$ p$_{\rm T,ch. jet}$ are above the simulation at small $z_{\rm ||}^{\rm ch,jet}$ and below for larger values of momentum fraction. The simulation provides a better description of the data for the higher jet-momentum interval.

\begin{figure}[bth]
\centering
\begin{subfigure}[b]{0.48\textwidth}
\includegraphics[width=18pc]{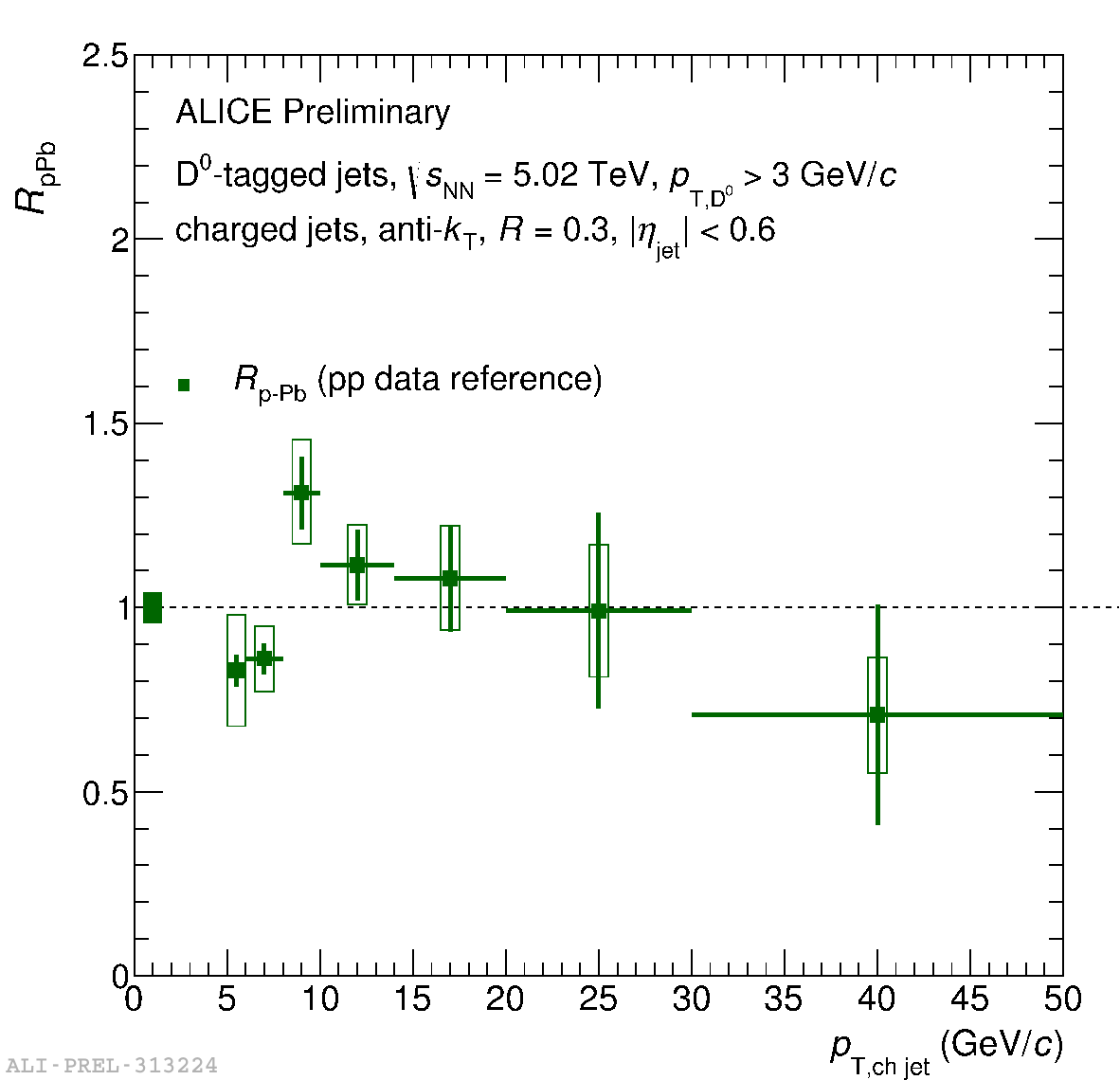}
\caption{$R_{\rm pPb}$.}
\label{fRpA_D0jet}
\end{subfigure}
\begin{subfigure}[b]{0.48\textwidth}
\includegraphics[width=18pc]{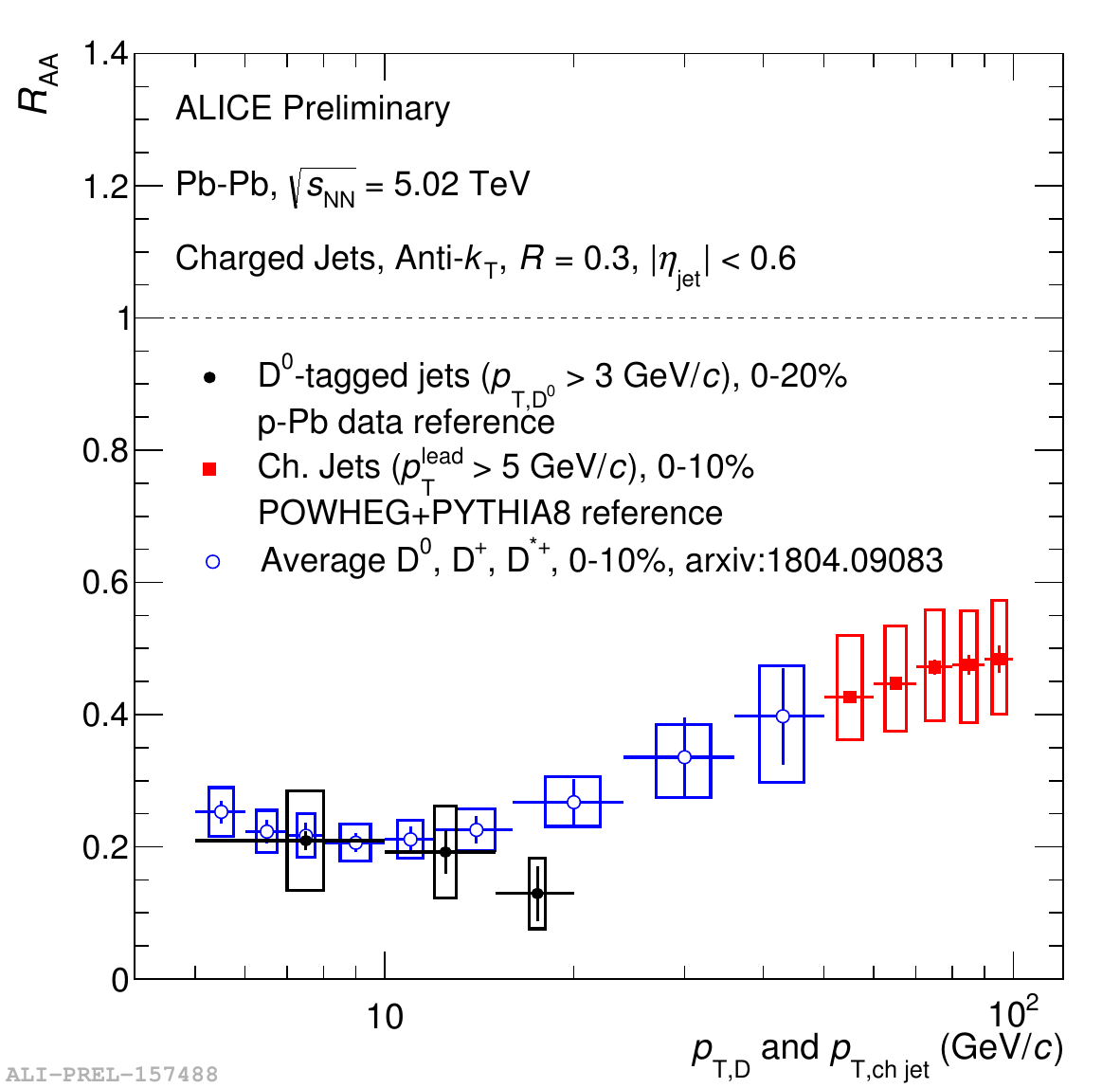}
\caption{$R_{\rm AA}$.}
\label{fRAA_D0jet}
\end{subfigure}
\caption{Nuclear modification factor in p--Pb ($R_{\rm pPb}$) and Pb--Pb ($R_{\rm AA}$) collisions at $\sqrt{s_{\rm NN}}=$ 5.02 TeV.}
\label{fHI_nmf}
\end{figure}

The nuclear modification factors (Fig. \ref{fHI_nmf}) were measured for D$^{0}$-tagged jets in p--Pb and Pb--Pb collisions. This observable is defined as the ratio of the $p_{\rm T}$-differential yields of jets in heavy-ion and pp collisions scaled by the Pb atomic mass, in case of p--Pb collisions,  or the number of binary collisions \cite{Glauber}, in case of Pb--Pb collisions. The $R_{\rm pPb}$ is shown in Fig. \ref{fRpA_D0jet} and it is consistent within uncertainties with unity, indicating that cold nuclear matter effects are small or negligible. Figure \ref{fRAA_D0jet} presents the $R_{\rm AA}$ of D$^{0}$-tagged jets (black markers). A strong suppression of D$^{0}$-tagged jets is observed in central (0-20\%) Pb--Pb collisions. The comparison with the $R_{\rm AA}$ of the D mesons (blue markers) indicates similar suppression up to 20 GeV/\textit{c}. At higher jet transverse momentum the $R_{\rm AA}$ of inclusive jets (red markers) and the average of D mesons show a similar trend. 

\subsection{Jets tagged by electrons from heavy-flavour decays}
Jets containing an electron from heavy-flavour hadron decays (charm and bottom) were reconstructed using anti-$k_{\rm t}$ algorithm with resolution parameter $R$ = 0.3. An invariant mass analysis of e$^{+}$e$^{-}$ pairs is performed to remove the contribution of "photonic" electrons from gamma conversion and $\pi^{0}$ and $\eta$ Dalitz decays. Pairs with invariant mass smaller than 0.1 GeV/\textit{c}$^{2}$ are labeled as photonic electrons and removed from the sample. Ultimately, an unfolding procedure is performed to correct for detector effects. In p--Pb collisions unfolding also takes into account the fluctuations of the underlying event energy density.

\begin{figure}[bth]
\centering
\begin{subfigure}[b]{0.48\textwidth}
\includegraphics[width=18pc]{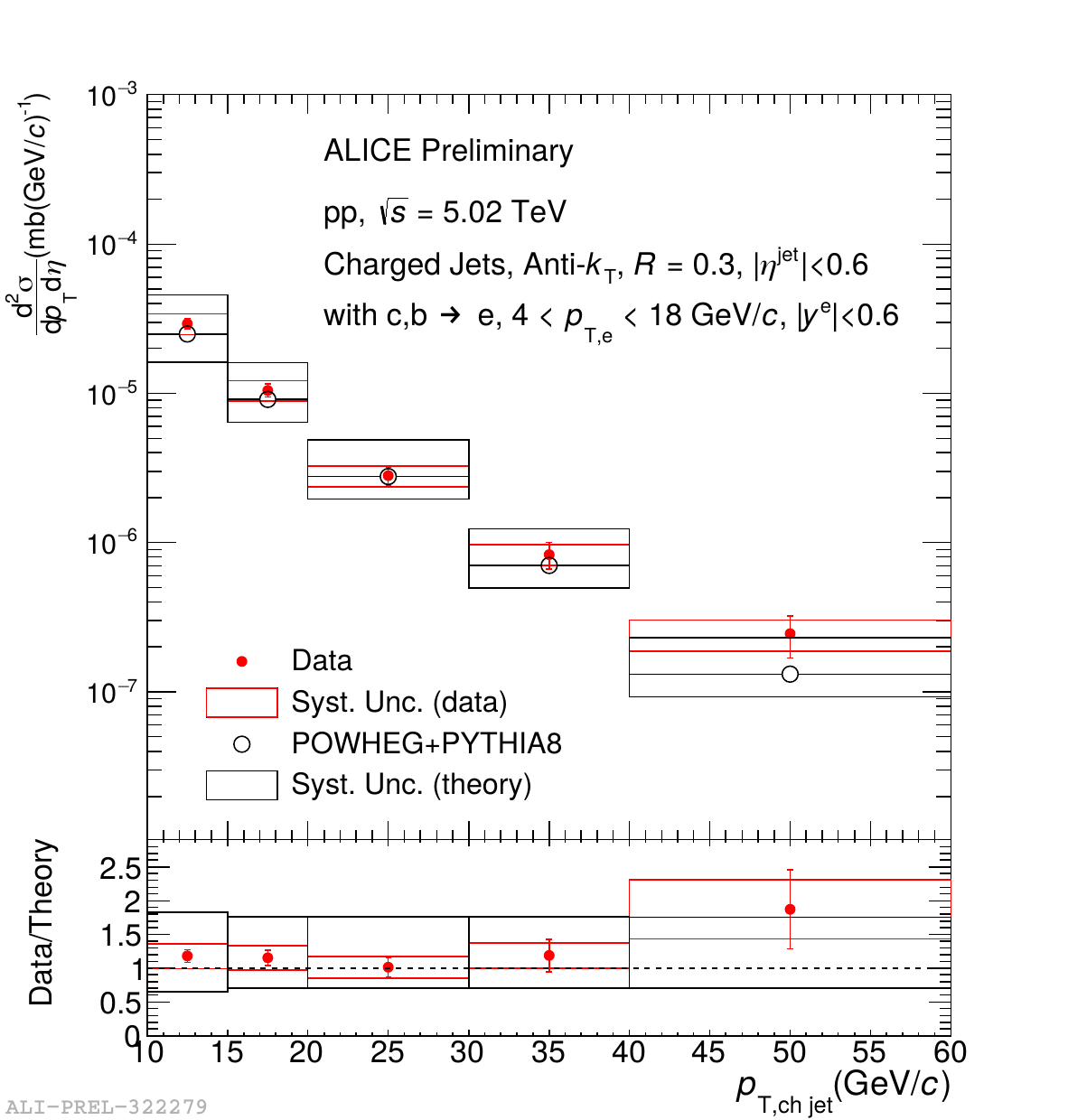}
\caption{pp collisions.}
\label{fHFe_pp}
\end{subfigure}
\begin{subfigure}[b]{0.48\textwidth}
\includegraphics[width=18pc]{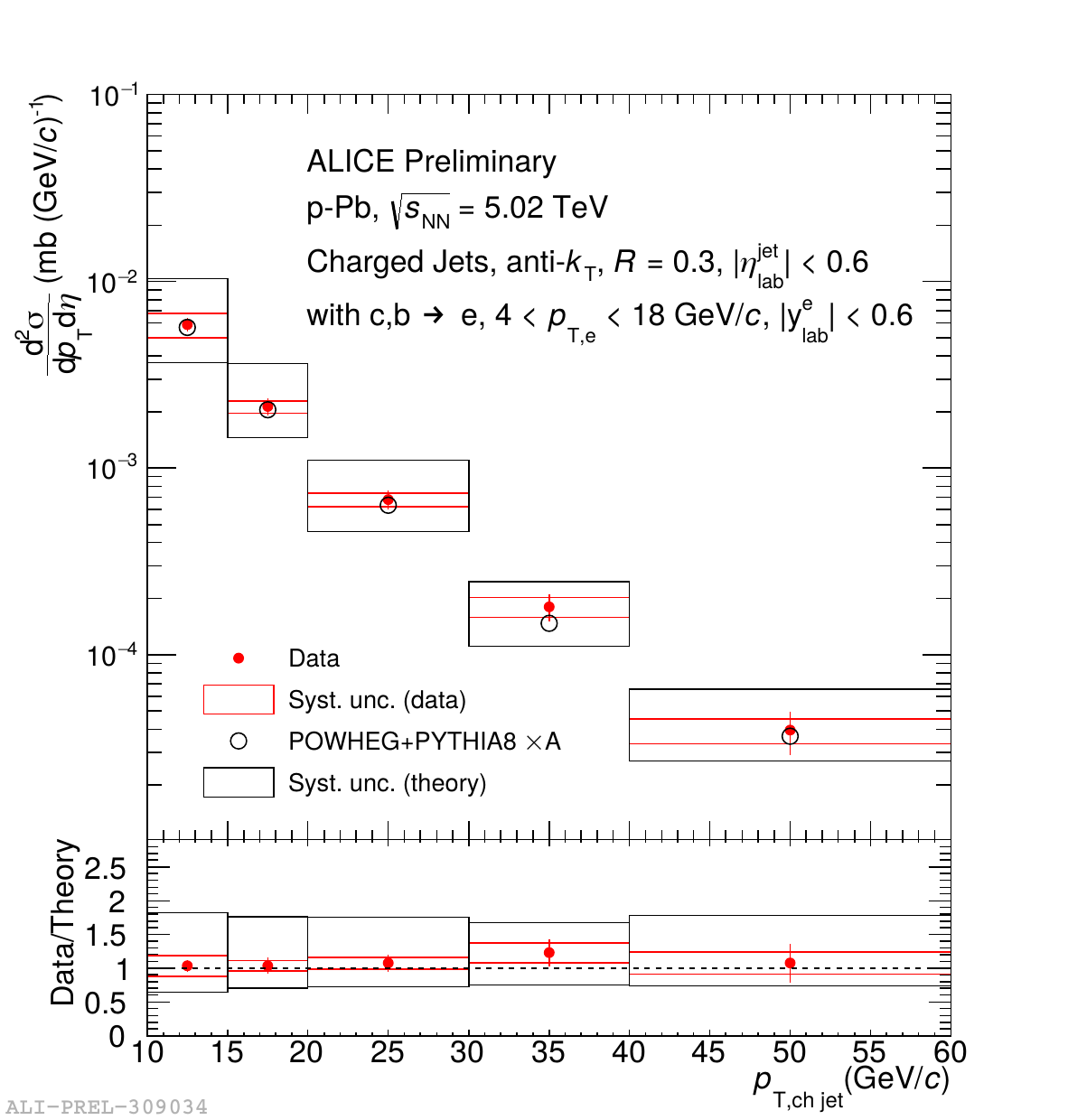}
\caption{p--Pb collisions.}
\label{fHFe_pPb}
\end{subfigure}
\caption{$p_{\rm T}$-differential cross section of charm and bottom jets tagged with electrons from heavy-flavour decays in pp and p--Pb collisions at $\sqrt{s_{\rm NN}}=$ 5.02 TeV.}
\label{fHFe_JetSpectra}
\end{figure}

The transverse momentum dependent spectra of jets tagged with electrons from heavy-flavour decays are presented in Fig. \ref{fHFe_JetSpectra} in pp (Fig. \ref{fHFe_pp}) and p--Pb (Fig. \ref{fHFe_pPb}) collisions at $\sqrt{s_{\rm NN}}=$ 5.02 TeV. A good description of the data is provided by POWHEG+PYTHIA8 simulations.

\begin{figure}[bth]
\centering
\includegraphics[width=24pc]{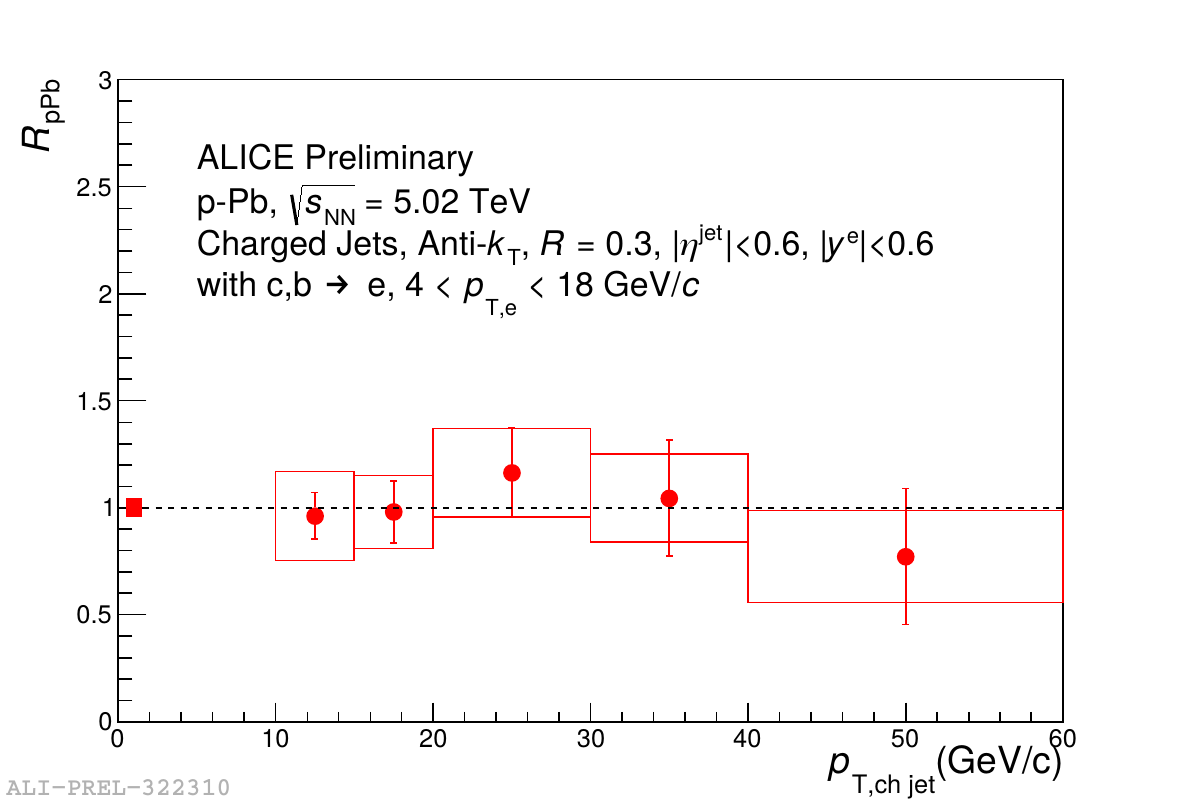}
\caption{Nuclear modification factor ($R_{\rm pPb}$) of heavy-flavour jets tagged by electrons from heavy-flavour decays in p--Pb collisions at $\sqrt{s_{\rm NN}}=$ 5.02 TeV.}
\label{fHFe_RpPb}
\end{figure}

The measured $R_{\rm pPb}$ shown in Fig. \ref{fHFe_RpPb} is compatible with unity, indicating no significant modifications from cold nuclear matter effects.

\subsection{b-jets}

Jets from bottom quark fragmentation are identified by the presence of a decay vertex, displaced from the collision point, reconstructed with 3 particles within the constituents of the jet. A selection exploiting the relatively long decay length of hadrons with bottom quarks is applied to enhance the discriminating power of the b-tagging. This selection is done mainly using the significance $L_{\rm xy}/\sigma_{L_{\rm xy}}$, where $L_{\rm xy}$ is the distance between the primary and secondary vertices in the xy-plane and $\sigma_{L_{\rm xy}}$ is its uncertainty. The variable $\sigma_{\rm SV}=\sqrt{\sum_{i=1}^{3} d_{i}^{2}}$, where $d$ is the distance of closest approach between each of the three tracks and the position of the secondary vertex is also powerful to identify jets from bottom quarks. A resolution parameter $R$ = 0.4 was used for jet finding.

\begin{figure}[bth]
\centering
\includegraphics[width=24pc]{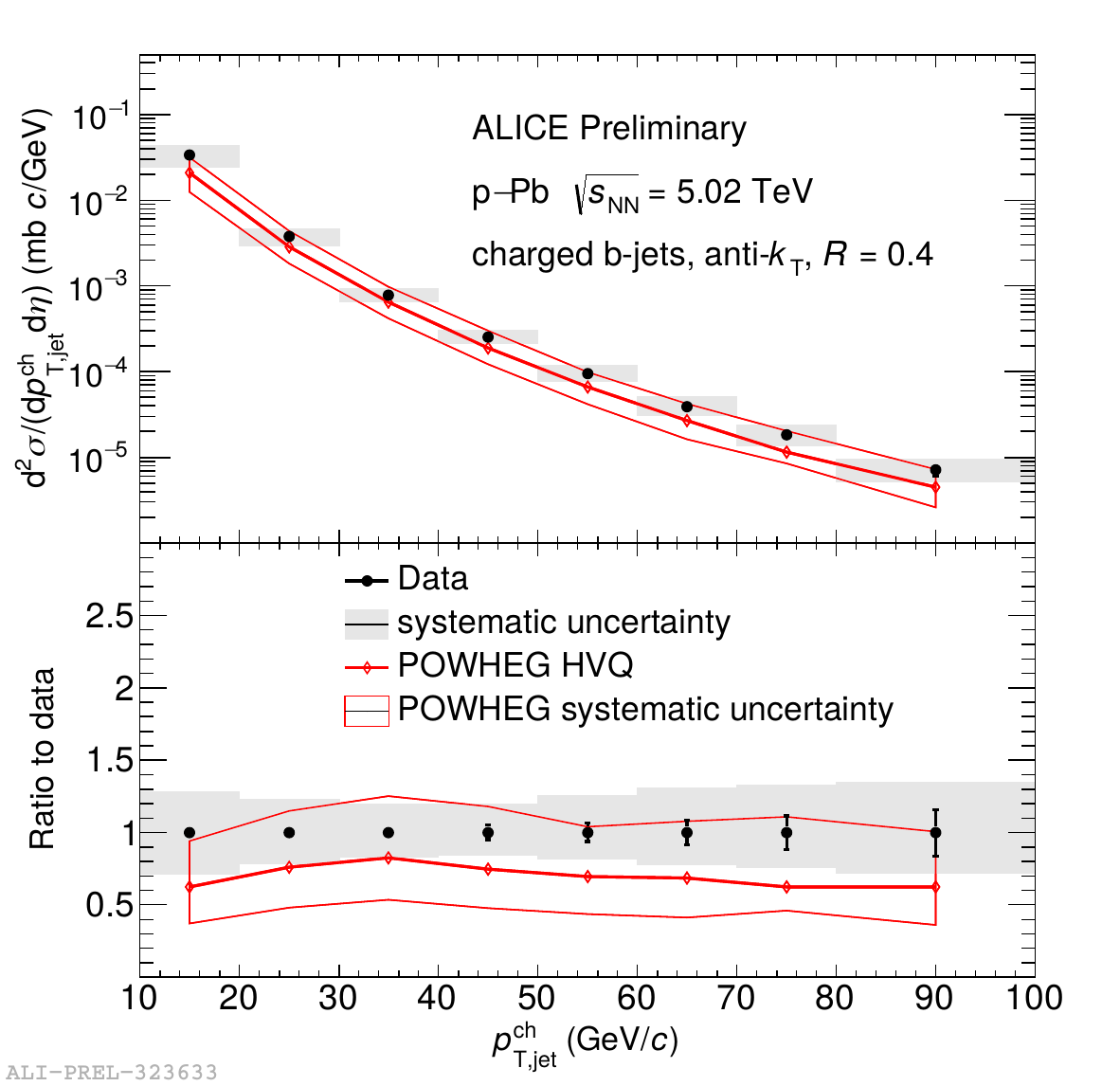}
\caption{$p_{\rm T}$-differential cross section of b-jets in p--Pb collisions at $\sqrt{s_{\rm NN}}=$ 5.02 TeV.}
\label{fbjets}
\end{figure}

The $p_{\rm T}$-differential cross section of jets from bottom quarks is presented in Fig. \ref{fbjets}. The tagging efficiency was estimated with a Monte Carlo simulation in which a PYTHIA event containing a b-$\bar{\rm b}$	pair was injected on top of a p--Pb collision simulated with EPOS \cite{PhysRevC.92.034906}. The purity was calculated using POWHEG simulations and has a approximate value of 0.4, but varying as a function of jet $p_{\rm T}$. Detector finite resolution and jet background fluctuations are corrected for with a SVD unfolding. The data are described within uncertainties by POWHEG-HVQ\cite{Frixione:2007nu} simulations.

\subsection{D-hadron correlations}

The reconstruction of D mesons (D$^{0}$, D$^{+}$ and D$^{*+}$) is performed as in subsection \ref{DJets} and are correlated with primary charged particles. The angular correlation distribution is corrected for the finite acceptance of the ALICE detector using an event-mixing technique. The D-meson combinatorial background is removed via an invariant mass analysis. The feed-down contribution from bottom quark decays is removed using FONLL calculations \cite{Cacciari:1998it} and templates of the correlation function obtained from simulations performed with PYTHIA event generator. The D-hadron azimuthal correlation distribution is fitted using two gaussian functions for the near and away side peaks and a constant function for the baseline \cite{Acharya:2019icl}.

\begin{figure}[bth]
\centering
\includegraphics[width=26pc]{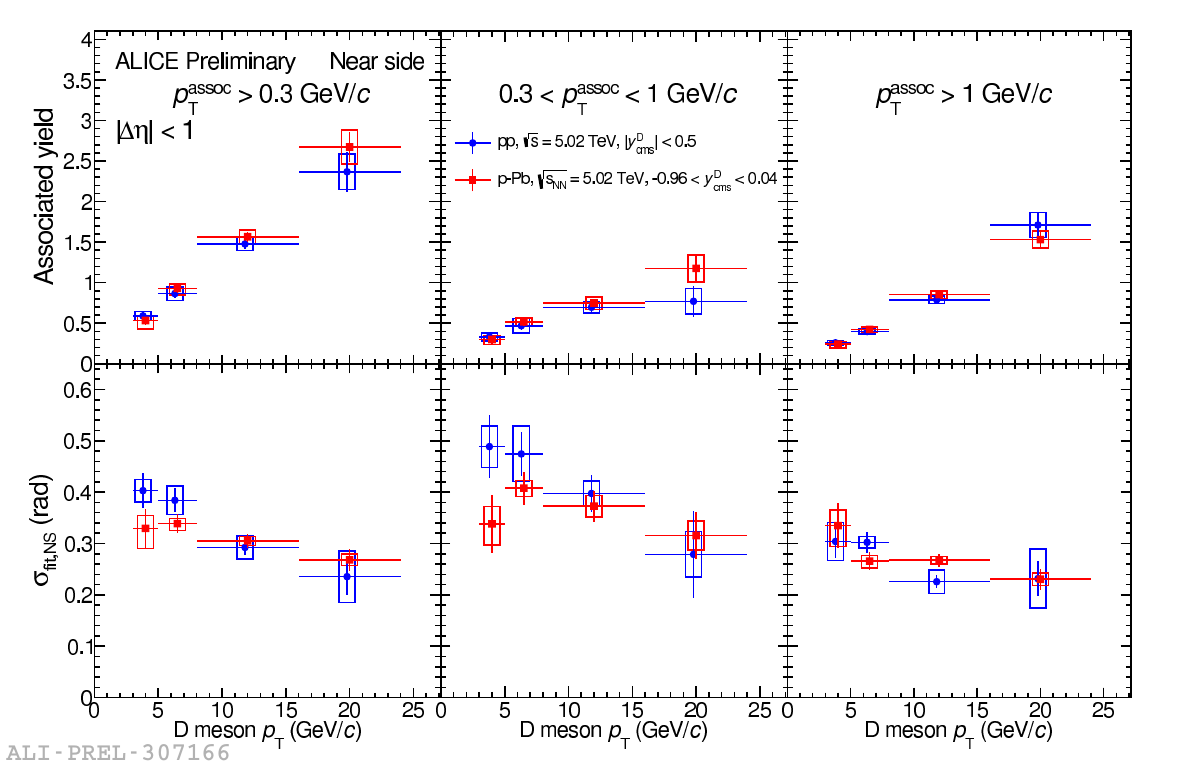}
\caption{Associated yields and standard deviation of the near-side peak obtained from D-hadron angular correlations in pp and p--Pb collisions at $\sqrt{s_{\rm NN}}=$ 5.02 Tev.}
\label{fDHadronCorr}
\end{figure}

The associated yields and the standard deviation ($\sigma_{\rm fit,NS}$) of the Gaussian function of the near side peak as a function of D-meson transverse momentum are presented in Fig. \ref{fDHadronCorr} for pp and p--Pb collisions at $\sqrt{s_{\rm NN}}=$ 5.02 TeV. The associated yields are studied in three intervals of charged particle transverse momentum. For all intervals an enhancement of the associated yield in the near side peak is observed for increasing D-meson transverse momentum ($p_{\rm T}$). This is an indication that the higher the D-meson $p_{\rm T}$, the higher is the number of particles close to the D-meson in the $\eta-\phi$ phase space, which provides complementary piece of information to the change of the momentum fraction for large jet transverse momentum presented in subsection \ref{DJets}.
The standard deviation of the near-side Gaussian function also shows that particles in the heavy-flavour jet are more collimated for large D-meson transverse momentum.
No significant difference between pp and p--Pb measurements were observed in the associated yield or standard deviation of the near-side Gaussian function, indicating that they are not sensible to cold nuclear matter effects.

\section{Summary}
The probability distributions of the D$^{0}$-tagged jet momentum fraction in pp collisions at $\sqrt{s}=$ 5.02 and 13 TeV were presented. An indication of softer fragmentation of charm quarks at the lower jet momentum region is observed. The $R_{\rm pPb}$ of D$^{0}$-tagged jets indicate that cold nuclear matter effects are not significant. The $R_{\rm AA}$ indicates a strong suppression of D$^{0}$-tagged jet production in central (0--20\%) Pb--Pb collisions. The $R_{\rm pPb}$ of jets tagged with electrons from heavy-flavour decays was measured at $\sqrt{s_{\rm NN}}=$ 5.02 TeV and no significant cold nuclear matter effects were observed. The transverse momentum distribution of b-jets in p-Pb collisions at $\sqrt{s_{\rm NN}}=$ 5.02 TeV was reported in the kinematic range 10 $<$ jet $p_{\rm T}$ $<$ 100 GeV/$c$. D-hadron correlations studies in pp and p--Pb at $\sqrt{s_{\rm NN}}=$ 5.02 TeV were presented. No significant cold nuclear matter effects were observed.

\section*{References}

\bibliographystyle{unsrt}
\bibliography{Bibliography}
\end{document}